\documentclass[12pt]{article}
\usepackage{hyperref}

\usepackage[margin=1in]{geometry}
\usepackage{amsmath, amssymb}
\usepackage{booktabs}
\usepackage{array}
\usepackage{longtable}
\usepackage{makecell}
\usepackage{tikz}
\usepackage{setspace}
\usepackage{hyperref}

\title{\textbf{Optimal Networks with Accumulative Costs and Bidirectional Communication}}
\author{Juan M. C. Larrosa\thanks{Department of Economics, Universidad Nacional del Sur; Instituto de Ciencias e Ingeniería de la Computación (ICIC). E-mail: \texttt{jlarrosa@uns.edu.ar}.}
\and Fernando A. Tohm\'e\thanks{Department of Economics, Universidad Nacional del Sur; Instituto de Matemática de Bahía Blanca (INMABB). E-mail: \texttt{ftohme@uns.edu.ar}.}}
\date{}

\begin{document}

\maketitle

\begin{abstract}
This paper is an extension of the approach of Larrosa and Tohm\'e (2003), in which the payoff function is modified by allowing information to flow in both directions. Costs continue to be paid by the agent who initiates the connection, so this asymmetry reveals changes in the final equilibrium topology. We find that several optimal topologies persist as Nash networks, but that a strict Nash network corresponds to the sequential line network with intermediate active nodes; that is, agents are placed in a line and, every other agent, they connect with the predecessor agent and with the subsequent agent. In this way, cost accumulation is interrupted and benefits are maximized.


\end{abstract}

\section*{Introduction}

Interaction among agents can be represented in several ways. One way of representing direct exchanges that has attracted considerable interest in recent years is through networks. Since it is easy to understand, this analytical tool was first adopted in sociology and anthropology. For experts in those areas, it constitutes a pictorial way of understanding the influence of agents' neighborhoods on individual behavior. Focused on real social networks, sociologists and anthropologists have accumulated a vast amount of evidence that helps explain how human behavior is shaped by the behavior of other agents.

In mathematical terms, a network is a graph, where nodes represent individual agents and arcs or links represent the utility good, for example information, personal prestige, and so forth, that is exchanged. The economic literature has recently introduced game-theoretic tools into this analytical framework. Rather than being interested only in descriptive aspects, some economic theorists have addressed the study of how networks are formed in the first place, and what makes them stable or efficient; see Jackson and Wolinsky (1996), Bala and Goyal (2000), and Dutta and Jackson (2000). The network approach based on game theory exhibits two main strands: one based on cooperative games and another with a strong strategic sense. The analysis based on cooperative games, as is usual in this approach, studies the problem of coalition formation among agents. The strong assumption of utility transfer among agents is difficult to justify in many cases, as well as often computationally costly; see Qin (1996), Dutta et al. (1998), and Slikker and van den Nouweland (2001).

The strategic or noncooperative approach, in turn, only requires the definition of the strategies available to the agents, as well as the characterization of the corresponding payoff functions. Given a certain protocol or rule of interaction, agents decide whether or not to connect to the network, evaluating the benefits of connection or disconnection with other agents. The rational decisions of the agents lead to Nash equilibria, which support the networks that are the focus of this analytical framework. This is precisely the approach adopted in this paper.

This paper follows the line of Larrosa and Tohm\'e (2003), although it now expands the possibility of information circulation. We consider that information flows bidirectionally, while links are supported by those who initiate the links. Under this payoff scheme, the cost function is accumulative. As information passes through more agents, it becomes more expensive. What optimal communication topology supports these conditions? We find that a type of sequential linear network is the one that allows the minimum number of connections with the maximum flow of information.

Section I begins the analysis with the definition of bidirectional-flow graphs. Section II presents the model. Section III develops the equilibrium architecture, as well as the way in which equilibria satisfy some criteria of stability and optimality. Section IV discusses the analogies and differences with the unidirectional case and concludes the paper.

\section*{I. The Case of Bidirectional Information Flow}

Networks will be modeled as directed graphs with bidirectional flows. By establishing links with other agents, the agent may obtain their information, but the others automatically access the information held by her. The agent who initiates the link is the one who must pay for it. This asymmetry will modify the optimal structures with respect to the case considered when the information flow is unidirectional; see Larrosa and Tohm\'e (2003).

\subsection*{I.1 Definitions}

The following concepts and tools, which are common to the development of the entire argument of this paper, are defined. Let
\[
N=\{1,\ldots,n\}
\]
be a set of agents. To avoid trivial results, we shall always assume that
\[
n\geq 3.
\]
If $i$ and $j$ are two typical members of $N$, a link between them, without intermediaries, originated in $i$ and ending in $j$, will be represented as $ij$. The interpretation of $ij$ is that $i$ establishes a contact with $j$, allowing $i$ to access $j$'s information as well as her network of contacts.

Each agent $i\in N$ has some information of her own,
\[
v_i\in \mathbb{Z}_{+},
\]
that is, represented as a positive integer. By convention, we initially assume that the information of each agent is sufficiently valuable for it to be worthwhile to establish a link with her,\footnote{With this assumption we rule out the possibility of the empty network, that is, the case in which all agents remain isolated, as an equilibrium structure in our model.} that is,
\[
v_i>1.
\]
As mentioned, $i$ can access more information by forming links with other agents. Agents will try to maximize the utility of the information available to them while minimizing the cost of connection with other agents. To achieve this, they will be endowed with a set of strategies. Each strategy for $i\in N$ is a vector
\[
g_i=(g_{i,1},\ldots,g_{i,i-1},g_{i,i+1},\ldots,g_{i,n})
\]
of $(n-1)$ dimensions, where each $g_{i,j}$, for $j\neq i$, takes the value 0 or 1. This is interpreted as follows: $i$ establishes a direct link with $j$ if
\[
g_{i,j}=1,
\]
whereas if
\[
g_{i,j}=0,
\]
such a link does not exist. The set of all strategies is denoted by $G_i$. The analysis is restricted only to the cases of pure strategies, which implies that
\[
|G_i|=2^{n-1}.
\]
Finally,
\[
G=G_1\times\cdots\times G_n
\]
denotes the set of strategy profiles in the interaction among the agents in $N$.

There is understood to exist a path from $j$ to $i$ according to $g\in G$ if there exists a sequence of different agents, in order to avoid the existence of cycles,
\[
j_0,\ldots,j_m,
\]
with
\[
i=j_0
\quad \text{and} \quad
j=j_m,
\]
such that
\[
g_{j_0,j_1}=\cdots=g_{j_{m-1},j_m}=1.
\]
In words, given a joint strategy $g$, such that
\[
j_1\in N^{g_{j_0}},\quad
j_2\in N^{g_{j_1}},\quad
\ldots,\quad
j_m\in N^{g_{j_{m-1}}}.
\]
A path from $j=j_m$ to $i=j_0$, denoted as
\[
j\to_g i,
\]
has a length equal to the cardinality of the sequence
\[
j_1,j_2,\ldots,j_{m-1},j_m,
\]
that is, $m$, which indicates the number of intermediate links between $j$ and $i$. Notice that a directed link is a path of length 1.

Other assumptions of the model are:

\begin{enumerate}
\item The value $v_i$ does not change as more players come to know that information.
\item The value of $v_i$ does not change as a particular individual comes to know more information.
\item Information is clearly transmitted through the network, with no distortion of information and no decay of flow, such that the value of $v_i$ for $j$ is the same whether $i$ and $j$ are connected directly or indirectly.
\end{enumerate}

The case in which information flows in two directions once contact has been established is analyzed below.

\section*{II. The Model}

This section also considers the case of directed graphs. Although information flows in both directions through a link, that is, bidirectional communication analogous to that observed in undirected graphs, these graphs are directed because they take into account the cost of initiating the link. Thus, $g_{ij}=1$ implies that $i$ finances the link and, therefore, does not have the same interpretation as $g_{ji}=1$, as required by the definition of an undirected graph. What is emphasized here is the symmetry in the benefits of those who are connected, since both enjoy the same information, and the asymmetry of costs, since whoever initiates the link must finance the connection.

The problem is to determine which structure can emerge as a strategic equilibrium among the agents and whether this structure is optimal. It is found that strict Nash equilibria, or Nash equilibria with the minimum number of links, support a peripherally financed star network, which is stable and optimal.

Consider an example analogous to Example 1 for the undirected case.

\textbf{Example 1.} Consider a group of four agents,
\[
N=\{a,b,c,d\}.
\]
A joint strategy
\[
g=(g_a,g_b,g_c,g_d)
\]
can be represented as a strategy profile. This can be seen in Table 1, where the asterisk indicates who initiated the link.

Each row is the strategy chosen by one of the agents. The columns correspond to the agents. An entry 1 in row $i$ and column $j$ means that the strategy of agent $i$ prescribes establishing a link with agent $j$. Entries on the main diagonal are marked with crosses, since agents cannot establish a link with themselves. Figure 1 shows the directed graph corresponding to $g$.

\begin{table}[h!]
\centering
\caption{Strategy profile for Example 1}
\begin{tabular}{c|cccc}
\toprule
Strategy & $a$ & $b$ & $c$ & $d$ \\
\midrule
$g_a$ & X & $1^*$ & 0 & 0 \\
$g_b$ & 1 & X & $1^*$ & 0 \\
$g_c$ & 0 & 1 & X & $1^*$ \\
$g_d$ & 0 & 0 & 1 & X \\
\bottomrule
\end{tabular}
\end{table}

\begin{figure}[h!]
\centering
\begin{tikzpicture}[>=stealth, node distance=2.5cm]
\node (a) at (0,2) {$a$};
\node (b) at (0,0) {$b$};
\node (c) at (3,0) {$c$};
\node (d) at (3,2) {$d$};

\draw[-] (b) -- (a);
\draw[-] (b) -- (c);
\draw[-] (c) -- (d);

\node[circle,draw,fill=white,inner sep=1.2pt] at (0,1.35) {};
\node[circle,draw,fill=white,inner sep=1.2pt] at (1.05,0) {};
\node[circle,draw,fill=white,inner sep=1.2pt] at (3,1.35) {};
\end{tikzpicture}
\caption{Network formed by the strategy profile of Example 1}
\end{figure}
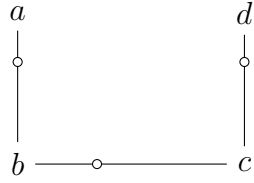

Following Bala and Goyal (2000, pp. 1190--1191), the closure of $g$ is denoted as
\[
\bar g=cl(g)
\]
and defined by
\[
\bar g_{i,j}=\max\{g_{i,j},g_{j,i}\}
\]
for each $i$ and $j$ in $N$. Define
\[
N_i^d(g)=\{k\in N\mid g_{i,k}=1\}
\]
as the set of agents with whom $i$ establishes a direct link according to her strategy profile $g_i$. Also denote by
\[
N_i^u(g)=\{h\neq k\mid h\in N,\ d_g(i,h)\geq 2\}
\]
the set of agents with whom $i$ establishes an indirect link according to her strategy profile $g_i$. Consider
\[
N^{i;g}=\{k\in N\mid i\leftrightarrow_g k\}
\]
as the set of agents with whom $i$ establishes a bidirectional directed link according to her strategy profile $g_i$.

There is a path from $j$ to $i$ according to $g\in G$ if there exists a sequence of different agents, in order to avoid cycles,
\[
j_0,\ldots,j_m,
\]
with
\[
i=j_0
\quad \text{and} \quad
j=j_m,
\]
such that
\[
g_{j_0,j_1}=\cdots=g_{j_{m-1},j_m}=1.
\]
In words, given a joint strategy $g$, we have
\[
j_1\in N^{g_{j_0}},\quad
j_2\in N^{g_{j_1}},\quad
\ldots,\quad
j_m\in N^{g_{j_{m-1}}}.
\]
A path from $j=j_m$ to $i=j_0$, denoted as
\[
j\leftrightarrow_g i,
\]
has a length equal to the cardinality of the sequence
\[
j_1,j_2,\ldots,j_{m-1},j_m,
\]
that is, $m$, which indicates the number of intermediate links between $j$ and $i$. Notice that a directed link is a path of length 1.

The existence of a direct link $ij$ indicates symmetric communication between $i$ and $j$. That is,
\[
g_{i,j}=1
\]
indicates that $i$ has established communication with $j$, which allows $i$ to access $j$'s information, but it also allows $j$ to access the information held by $i$. The asymmetry arises in costs. Here, $i$ has initiated the connection, and therefore she is the one who must finance the cost of the connection, while $j$ incurs no cost. Structures with this characteristic are called bidirectional-flow networks.

In bidirectional-flow networks, the strategy profile can be represented as a directed graph
\[
g=(g_1,\ldots,g_n)
\]
on $N$. That is, in the bidirectional directed graph, the elements of $N$ are the nodes, while each link established as
\[
g_{i,j}=1
\]
is represented by a line beginning at $j$ with a white circle on the line as it approaches $i$. This represents the idea that $i$ was the one who initiated the link, while information circulates in both directions.

The set of agents accessed, directly or otherwise, by $i$ is expressed as
\[
N^{i;g}=\{k\in N\mid k\to_g i\}.
\]
The agent $i$ is not included in order to facilitate the subsequent derivations of the model.\footnote{Strictly speaking, it should be
\[
N^{i;g}=\{k\in N\mid k\to_g i\}\cup\{i\}.
\]
The agent $i$ is included in $N^{i;g}$ to indicate that $i$ knows her own valuation, despite the previously mentioned fact that $i$ does not establish a direct link with herself.}
Notice that
\[
N^{i;g}=N_i^d(g)+N_i^u(g),
\]
in accordance with the preceding definitions.

Let
\[
\mu_i:G\to \{0,\ldots,n(n-1)\}
\]
be the number of links in all the paths that end in $i$, originated by agents in $N^{i;g}$ under any joint strategy:
\[
\mu_i(g)=
\left|
\left\{
(j,k)\in N\times N:
g_{j,k}=1,\ 
\exists l\in N^{i;g}\ \text{with}\ l\to_g i
\ \text{and}\ j,k\in l\to_g i
\right\}
\right|.
\]
Notice that there may be more than one path from $j$ to $i$.

\textbf{Example 2.} Suppose
\[
N=\{1,2,3,4,5\}
\]
and the strategy
\[
g=(g_1,g_2,g_3,g_4,g_5)
\]
is given by Table 2, where the asterisk indicates who initiated the connection.

\begin{table}[h!]
\centering
\caption{Strategy profile for Example 2}
\begin{tabular}{c|ccccc}
\toprule
Strategy & 1 & 2 & 3 & 4 & 5 \\
\midrule
$g_1$ & X & $1^*$ & 0 & 0 & $1^*$ \\
$g_2$ & 0 & X & $1^*$ & 0 & 0 \\
$g_3$ & 0 & 0 & X & $1^*$ & $1^*$ \\
$g_4$ & 0 & 0 & 0 & X & 0 \\
$g_5$ & 0 & 0 & 0 & 0 & X \\
\bottomrule
\end{tabular}
\end{table}

\begin{figure}[h!]
\centering
\begin{tikzpicture}[>=stealth]
\node (one) at (0,2) {1};
\node (two) at (0,0) {2};
\node (three) at (3,0) {3};
\node (four) at (6,0) {4};
\node (five) at (3,2) {5};

\draw[-] (two) -- (one);
\draw[-] (three) -- (two);
\draw[-] (four) -- (three);
\draw[-] (five) -- (three);
\draw[-] (five) -- (one);

\node[circle,draw,fill=white,inner sep=1.2pt] at (0,1.3) {};
\node[circle,draw,fill=white,inner sep=1.2pt] at (1.0,0) {};
\node[circle,draw,fill=white,inner sep=1.2pt] at (4.8,0) {};
\node[circle,draw,fill=white,inner sep=1.2pt] at (3,1.35) {};
\node[circle,draw,fill=white,inner sep=1.2pt] at (1.7,2) {};
\end{tikzpicture}
\caption{Network formed by the strategy profile of Example 2}
\end{figure}
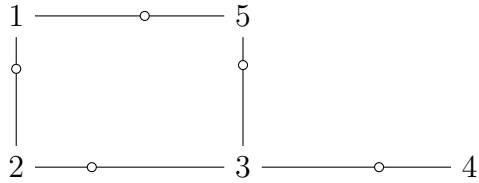

Figure 2 shows the corresponding network. We have
\[
N^{1;g}=\{1,2,3,4,5\},\quad
N^{2;g}=\{1,2,3,4,5\},
\]
\[
N^{3;g}=\{1,2,3,4,5\},\quad
N^{4;g}=\{1,2,3,4,5\},
\]
and
\[
N^{5;g}=\{1,2,3,4,5\}.
\]
Notice that under unidirectional information flow the results would be
\[
N^{1;g}=\{1,2,3,4,5\},\quad
N^{2;g}=\{2,3,4,5\},
\]
\[
N^{3;g}=\{3,4,5\},\quad
N^{4;g}=\{4\},
\quad
N^{5;g}=\{5\}.
\]
That is, under unidirectional information in $g$, agent 1 accesses the information of all agents, whereas 4 and 5 only access their own information. Now, with bidirectional information, under $g$ everyone accesses everyone's information. The numbers of links required to obtain the information are
\[
\mu_1(g)=5,\quad
\mu_2(g)=3,\quad
\mu_3(g)=2,
\]
whereas
\[
\mu_4(g)=\mu_5(g)=0.
\]
This does not change for mono- or bidirectional flow.

\subsection*{II.1 Connections and Benefits}

For the present analysis in particular, it will be fundamental to model the forms of connection that will prevail in the future determination of the payoff and cost functions of each agent. In the model, it will be crucial to distinguish the agent's direct links from her indirect links. Direct links are those links that have a path of value one, whereas indirect links are those that have a path of value greater than one, as already defined in the previous section:
\begin{equation}
N^{i;g}=
\{k\in N\mid g_{i,k}=1\}
+
\{h\neq k\mid h\in N,\ d_g(i,h)\geq 2\}.
\tag{1}
\end{equation}

Again, $N^{i;g}$ is a vector that collects all those agents who are directly or indirectly connected with $i$. Here, the degree of proximity of the links is also discriminated. The first term in (1) corresponds to the direct links of agent $i$, while the second term refers to the indirect links of the agent, which, as we shall also see, she must finance. That is, the payoffs of $i$ are the sum of all the information that can be accessed by her, minus the cost of the paths, that is, direct and indirect connections, that reach her and are established according to $g$. Recall that each link is assumed to have unit cost. The intuition here is that $i$ obtains a payoff from accessing more information, but at the same time she must pay a charge or fee for each of the links in the paths toward the sources of information. This way of financing connections is therefore accumulative. As information passes through more agents, it becomes more expensive, since all intermediaries must be paid.

Figure 3 illustrates that $i$ connects directly with $j_1$ and $j_2$, from whom she obtains information directly. But she also obtains indirect information because $j_2$ is connected with $k_1$, while $k_1$ is connected with $k_2$, and $k_2$ with $k_3$. Ultimately, $i$ accesses the information of both sets and must pay for the links included within both sets. Notice that $j_3$, $j_4$, and $k_4$ are never accessed by agent $i$.

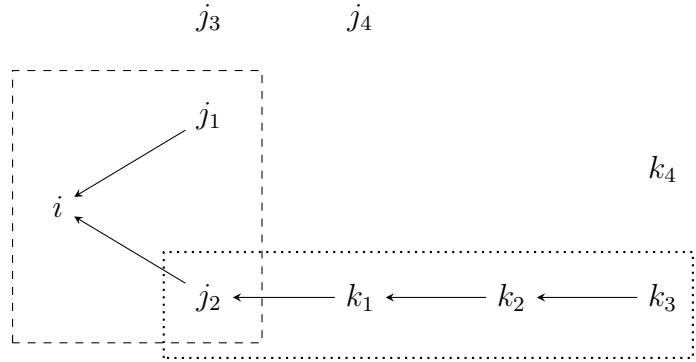
\begin{figure}[h!]
\centering
\begin{tikzpicture}[>=stealth]
\node (i) at (0,0) {$i$};
\node (j1) at (2,1.2) {$j_1$};
\node (j2) at (2,-1.2) {$j_2$};
\node (k1) at (4,-1.2) {$k_1$};
\node (k2) at (6,-1.2) {$k_2$};
\node (k3) at (8,-1.2) {$k_3$};
\node (j3) at (2,2.5) {$j_3$};
\node (j4) at (4,2.5) {$j_4$};
\node (k4) at (8,0.5) {$k_4$};

\draw[->] (j1) -- (i);
\draw[->] (j2) -- (i);
\draw[->] (k1) -- (j2);
\draw[->] (k2) -- (k1);
\draw[->] (k3) -- (k2);

\draw[dashed] (-0.6,-1.8) rectangle (2.7,1.8);
\draw[dotted, thick] (1.4,-2.0) rectangle (8.4,-0.6);
\end{tikzpicture}
\caption{Scheme of total connections and information accessed under unidirectional flow}
\end{figure}

The connection function stipulated above remains in force for this case, but it acquires a larger dimension given the assumption that information circulates in both directions. A graphical illustration is useful for this purpose. In Figure 3, the connection function is presented for the unidirectional case. In that case, the two accessed sets accounted for the cost structure and the information accessed by agent $i$. In other words, the dotted sets coincided insofar as within them, direct and indirect connections, were the costs that agent $i$ had to pay, but also within them were the revenues from accessing more information. However, now the cost structure remains associated with the same dotted sets of the unidirectional case, while the payment for information is the information held by all members of the network, since each piece of information is accessed merely through the existence of some type of connection between two agents, regardless of who initiates the link.

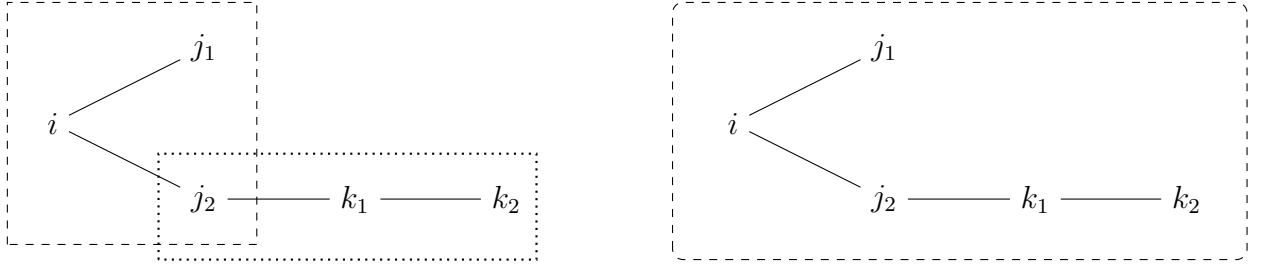
\begin{figure}[h!]
\centering
\begin{tikzpicture}[>=stealth]
\node at (0,2.2) {(a) Cost structure};
\node (i) at (0,0) {$i$};
\node (j1) at (2,1) {$j_1$};
\node (j2) at (2,-1) {$j_2$};
\node (k1) at (4,-1) {$k_1$};
\node (k2) at (6,-1) {$k_2$};

\draw[-] (i) -- (j1);
\draw[-] (i) -- (j2);
\draw[-] (j2) -- (k1);
\draw[-] (k1) -- (k2);
\draw[dashed] (-0.6,-1.6) rectangle (2.7,1.6);
\draw[dotted, thick] (1.4,-1.8) rectangle (6.4,-0.4);

\begin{scope}[xshift=9cm]
\node at (0,2.2) {(b) Accessed information};
\node (a) at (0,0) {$i$};
\node (b) at (2,1) {$j_1$};
\node (c) at (2,-1) {$j_2$};
\node (d) at (4,-1) {$k_1$};
\node (e) at (6,-1) {$k_2$};

\draw[-] (a) -- (b);
\draw[-] (a) -- (c);
\draw[-] (c) -- (d);
\draw[-] (d) -- (e);
\draw[rounded corners, dashed] (-0.8,-1.8) rectangle (6.8,1.6);
\end{scope}
\end{tikzpicture}
\caption{Connection structure and accessed information under bidirectional flow}
\end{figure}

The formal expression of each agent's payoff function is as follows. She has her own information, $v_i$, but she can access more information if she creates a connection function and accesses the information owned and obtained by other agents. Therefore, formalizing the value of the total information accessed by $i$, $I_i$, we obtain:
\begin{equation}
I_i(g)=v_i+\sum_{\substack{j\neq i\\ j\in N^{i;g}}} I_j(g),
\tag{2}
\end{equation}
where $I_j$ is the value of the information of agent $j$ that is accessed by $i$ through the strategy profile $g$, represented by $N^{i;g}$. This determines the connection structure of agent $i$ and the diversified information to which she has access.

To make this scheme a game, we must define the agents' benefits. We shall assume that
\[
\Pi_i:G\to \mathbb{R},
\]
the benefit function for agent $i$, is:
\begin{equation}
\Pi_i(g)\equiv
\sum_{j\in N^{i;g}} I_j(g)-c\,\mu_i(g),
\tag{3}
\end{equation}
where $I_i$ represents the information held and accessed by agent $i$ according to the connection strategy $g$ in definition (1), $c$ represents the cost of each connection, which for our particular case is assumed to be $c=1$, and $\mu_i(g)$, as defined in the previous section, is the number of direct and indirect links that agent $i$ must finance.

Consider now an important proposition.

\textbf{Proposition 1.} Given two joint strategies $g$ and $g'$, $\Pi_i(g)\geq \Pi_i(g')$ if and only if the corresponding graphs $N^{i;g}$ and $N^{i;g'}$ are such that:
\[
I_i(g)-I_i(g')\geq \mu_i(g)-\mu_i(g').
\]

\textit{Proof.} Trivial. \hfill $\square$

This result helps to understand the intuition that the objective of the rational agent is to obtain as much information as possible while crossing the smallest possible number of links. Two cases are of particular interest:
\[
\sum_{j\in N^{i;g}} I_j
=
\sum_{j\in N^{i;g'}} I_j
\quad \text{and} \quad
\mu_i(g)\leq \mu_i(g'),
\]
\[
\sum_{j\in N^{i;g}} I_j
\geq
\sum_{j\in N^{i;g'}} I_j
\quad \text{and} \quad
\mu_i(g)=\mu_i(g').
\]

The first shows that $\Pi_i(g)\geq \Pi_i(g')$ if the information obtained through $g$ is the same as the information obtained through $g'$, but the number of required links is smaller in $g$ than in $g'$. The second case shows that $\Pi_i(g)\geq \Pi_i(g')$ if the number of links required to reach the information is the same in $g$ as in $g'$, but the amount of information obtained in $g$ is greater than the amount obtained in $g'$.

The equilibrium situation that agents now face is considered next.

\section*{III. Equilibrium and Optimality}

Given a network
\[
g\in G,
\]
let $g_{-i}$ be the directed graph obtained when all direct links of agent $i$ are removed. Then $g$ can be written as
\[
g=(g_i,g_{-i}),
\]
meaning that $g$ is formed by the union of the links in $g_i$ and those in $g_{-i}$. A strategy $g_i$ is said to be a best response of agent $i$ to $g_{-i}$ if
\begin{equation}
\Pi_i(g_i,g_{-i})\geq \Pi_i(g_i',g_{-i})
\tag{4}
\end{equation}
for every
\[
g_i'\in G_i.
\]

The set of best responses to $g_{-i}$ is
\[
BR_i(g_{-i}).
\]
A network
\[
g=(g_1,\ldots,g_n)
\]
is said to be a Nash network if, for each $i$,
\[
g_i\in BR_i(g_{-i}),
\]
that is, if $g$, as a joint strategy, is a Nash equilibrium. To determine the structure of Nash networks, we provide several definitions that will allow us to describe additional properties of networks.

Given a network $g$, a set
\[
C\subset N
\]
is called a component of $g$ if, for every pair of agents $i$ and $j$ in $C$, with $i\neq j$, we have
\[
j\in N^{i;g},
\]
and there is no $C'$, with
\[
C\subset C',
\]
for which this is true. A component $C$ is said to be minimal if $C$ ceases to be a component once
\[
g_{i,j}=1
\]
between two agents $i$ and $j$ in $C$ is cut, that is, if
\[
g_{i,j}=0.
\]

We shall also define the concept of an active and passive node or player: if $i$ forms a link with $j$, then $i$ is an active neighbor of $j$; if, additionally, $j$ does not form a link with $i$, then $j$ is a passive neighbor of $i$.

A network is said to be connected if it supports a single component. If that single component is minimal, $g$ is said to be minimally connected. A network that is not connected is said to be disconnected. A particular instance of minimally connected networks is the line network, denoted $g^l$ from now on, in which agents can be labeled, by means of a function
\[
\ell:N\to N,
\]
as
\[
\{\ell(1),\ldots,\ell(n)\}
\]
and
\[
g_{\ell(1),\ell(2)}
=
g_{\ell(2),\ell(3)}
=
\cdots
=
g_{\ell(n-1),\ell(n)}
=1,
\]
with no other links.

Since the flow is bidirectional, it may occur that within the line network agents initiate connections in many different ways and still allow the flow of information along the network. That is, within the line network there are different sequential connection activation graphs that yield different payoffs to the agents. The line network with ordered activation nodes\footnote{In some papers this topology appears as a chain network. See, for example, Johnson and Gilles (2003).} maintains a connection configuration in which the first connects with the second, the second with the third, and so on until completing the sequence, with the penultimate connecting with the last, and no further connections.

Some forms of activation of the line network can be emphasized. For example, an alternative configuration is the line network with odd active nodes, denoted $g^{li}$, which is a linear topology that can be represented by the condition that odd nodes initiate links with their immediate predecessors and successors, while even nodes passively accept, until completing the connection sequence. Formally, for odd interior labels $r$,
\[
g_{\ell(r),\ell(r-1)}=g_{\ell(r),\ell(r+1)}=1,
\]
with the corresponding endpoint adjustment depending on whether $n$ is even or odd.

Finally, an analogous structure is the line network with even active nodes, denoted $g^{lp}$, which is the one satisfying the configuration in which the nodes labeled as even initiate connections toward their immediately neighboring nodes, while the rest of the nodes act passively and do not themselves initiate the connection. Formally, for even interior labels $r$,
\[
g_{\ell(r),\ell(r-1)}=g_{\ell(r),\ell(r+1)}=1,
\]
again with the corresponding endpoint adjustment depending on whether $n$ is even or odd.

Figure 5 shows an example with five agents for each of the cases described in this section.

\begin{figure}[h!]
\centering
\begin{tikzpicture}[>=stealth]
\node at (2,1.1) {Ordered activation};
\foreach \x in {1,...,5} {
\node[circle,draw,inner sep=1.5pt] (a\x) at (\x,0) {\x};
}
\draw[-] (a1)--(a2);
\draw[-] (a2)--(a3);
\draw[-] (a3)--(a4);
\draw[-] (a4)--(a5);

\begin{scope}[yshift=-2.4cm]
\node at (2,1.1) {Odd active nodes};
\foreach \x in {1,...,5} {
\node[circle,draw,inner sep=1.5pt] (b\x) at (\x,0) {\x};
}
\draw[-] (b1)--(b2);
\draw[-] (b3)--(b2);
\draw[-] (b3)--(b4);
\draw[-] (b5)--(b4);
\end{scope}

\begin{scope}[yshift=-4.8cm]
\node at (2,1.1) {Even active nodes};
\foreach \x in {1,...,5} {
\node[circle,draw,inner sep=1.5pt] (c\x) at (\x,0) {\x};
}
\draw[-] (c2)--(c1);
\draw[-] (c2)--(c3);
\draw[-] (c4)--(c3);
\draw[-] (c4)--(c5);
\end{scope}
\end{tikzpicture}
\caption{Different topologies of sequential-connection line networks}
\end{figure}

Thus, with all these elements in hand and by Proposition 1, the following result can be established. All results in this section correspond to the game
\[
(N,G,\Pi),
\]
where
\[
\Pi=\Pi_1\times\cdots\times\Pi_n.
\]

\textbf{Lemma 1.} Given the payoff function in (4), if $g^*$ is a strict Nash network, then it is a sequential-connection line network with either even or odd active nodes.

\textit{Proof.} Let
\[
\Pi_i:G\to \mathbb{Z}
\]
for each
\[
i\in N
\]
and let $g^*\in G$ be a strict Nash equilibrium. Then, for each $i$ and each $g_i\in G_i$, it holds that
\begin{equation}
\Pi_i(g_i^*,g_{-i}^*)>\Pi_i(g_i,g_{-i}^*).
\tag{5}
\end{equation}

It is said that $g^*$ defines a sequential-connection line network. If so, it must be true for every agent $i$ that
\[
\Pi_i(g)=
\begin{cases}
\displaystyle \sum_{h\in N} I_h-2, & \text{if } i \text{ is active},\\[6pt]
\displaystyle \sum_{h\in N} I_h-1, & \text{if } i \text{ is the last agent},\\[6pt]
\displaystyle \sum_{h\in N} I_h, & \text{if } i \text{ is passive}.
\end{cases}
\]

In words: the maximum amount of information that can be reached in a nonsequential-connection line network is the sum of all the information held by all agents, while the number of links that would make the information available to any one of them is strictly positive but low for active nodes, almost half of those connected, and zero for passive nodes, the other almost half of those connected. This is observed as one of the most efficient results in social terms, while individually it is also the optimal corollary.

Let $n$ be even. Suppose, by contradiction, that $g^*$ is not a sequential-connection line network.\footnote{For the proof it does not necessarily matter whether it is a network with even or odd active nodes, since the same result is obtained with trivial modifications.} This means that for at least one agent $i$,
\[
\Pi_i(g)\neq
\begin{cases}
\displaystyle \sum_{h\in N} I_h-2, & \text{if } i \text{ is active},\\[6pt]
\displaystyle \sum_{h\in N} I_h-1, & \text{if } i \text{ is the last agent},\\[6pt]
\displaystyle \sum_{h\in N} I_h, & \text{if } i \text{ is passive}.
\end{cases}
\]

First consider the case of an active agent for whom there could exist a deviation strategy such that
\[
\Pi_i(g^*)>\sum_{h\in N} I_h-2.
\]
Access to more information cannot be improved, since currently $i$ accesses all available information in the network. Therefore, a strategy that yields a higher benefit would have to evaluate only the reduction in the number of connections. If $i$ opted to connect only with agent $k+1$, the predecessor, or only with $k-1$, the successor, she would cancel the connectivity of the network by disconnecting it into two subnetworks. With this, she would access a lower level of information than
\[
\sum_{h\in N} I_h,
\]
having saved only one unit of utility in connection costs. On the other hand, if $i$ decided to connect with other agents located at $k\pm j$, such that
\[
2\leq j\leq n-j,
\]
then she would leave isolated the intermediate agents $k\pm k'$, with $k'<j$. Again, she would lose access to information more valuable than the benefit of reducing the corresponding connection costs. This is a contradiction.

Second, analyze the case in which $i$ is the last agent. If so, the only way to deviate would be to cut her only link with the rest of the network. But she would save one unit while losing all information except her own. This is a contradiction.

Third, consider now the case of an agent $i$ such that $i$ is passive. Let us see whether there exists a strategy $g^*$ such that
\[
\Pi_i(g^*)>\sum_{h\in N} I_h.
\]
Since $i$ accesses all information without incurring any cost, there is no action by $i$ that would allow her to increase her benefit. This is a contradiction. \hfill $\square$

Notice that not every Nash network is linearly sequential.

\textbf{Example 3.} Suppose
\[
N=\{1,2,3,4,5\}
\]
and the strategy
\[
g=(g_1,g_2,g_3,g_4,g_5)
\]
is given by Table 3.

\begin{table}[h!]
\centering
\caption{Strategy profile for Example 3}
\begin{tabular}{c|ccccc}
\toprule
Strategy & 1 & 2 & 3 & 4 & 5 \\
\midrule
$g_1$ & X & $1^*$ & $1^*$ & $1^*$ & $1^*$ \\
$g_2$ & 0 & X & 0 & 0 & 0 \\
$g_3$ & 0 & 0 & X & 0 & 0 \\
$g_4$ & 0 & 0 & 0 & X & 0 \\
$g_5$ & 0 & 0 & 0 & 0 & X \\
\bottomrule
\end{tabular}
\end{table}

In this case, there is an agent, agent 1, whose strategy $g_1$ is to connect with all the others, while the rest make no connection. This type of connection structure is called a centrally supported star network, and the network activation graph can be seen in Figure 6.

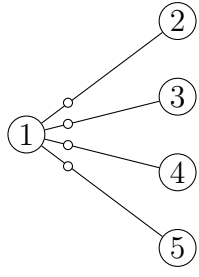
\begin{figure}[h!]
\centering
\begin{tikzpicture}[>=stealth]
\node[circle,draw,inner sep=1.5pt] (one) at (0,0) {1};
\node[circle,draw,inner sep=1.5pt] (two) at (2,1.5) {2};
\node[circle,draw,inner sep=1.5pt] (three) at (2,0.5) {3};
\node[circle,draw,inner sep=1.5pt] (four) at (2,-0.5) {4};
\node[circle,draw,inner sep=1.5pt] (five) at (2,-1.5) {5};

\draw[-] (one)--(two);
\draw[-] (one)--(three);
\draw[-] (one)--(four);
\draw[-] (one)--(five);

\node[circle,draw,fill=white,inner sep=1.2pt] at (0.55,0.42) {};
\node[circle,draw,fill=white,inner sep=1.2pt] at (0.55,0.14) {};
\node[circle,draw,fill=white,inner sep=1.2pt] at (0.55,-0.14) {};
\node[circle,draw,fill=white,inner sep=1.2pt] at (0.55,-0.42) {};
\end{tikzpicture}
\caption{Centrally supported star network}
\end{figure}

If analyzed, this network is a Nash network. Any deviation implies losses for the agent who incurs it. It is also efficient in the Pareto sense. However, this network presents an enormous asymmetry: a single active agent supports the entire network while the rest behave passively and incur no costs. The central agent has benefits
\[
\Pi_1(g)=\sum_{i\in N} I_i-(n-1),
\]
whereas the rest pay nothing and do receive all the information. Any structure that provides the same amount of information at lower costs can make agent 1 find deviation attractive, in accordance with Proposition 1.

Falk and Kosfeld (2003) find in network formation experiments that, for the case of the Bala and Goyal (2000) model, where costs are not accumulative but only direct-link costs are paid, the topology of the centrally supported star network never emerged as an equilibrium with experimental subjects. The authors emphasize that the high payoff asymmetry imposed on the central agent, as well as a larger general coordination problem, makes it unfeasible for any experimental subject to assume that role. Berninghaus et al. (2004) define this behavior as aversion to inequality, whose modeling requires considering certain qualitative aspects when forming links. They consider that active agents have greater information privileges than passive ones; that is, whoever creates a link accesses the agent and her network of contacts, while whoever allows herself to be accessed only accesses the agent. In this way, the authors manage to break the failure of an equilibrium structure to emerge in experiments with bidirectional information flow; in this case, the peripherally supported network appears as the equilibrium network.\footnote{The peripherally supported network is the inverse of the centrally supported network. In it, all agents except the central agent initiate links only with the central agent. The asymmetry compared with the previous case is notably reduced, since there are now $n-1$ active agents and only one passive agent.} This line of explanation supports discarding that topology as an equilibrium also in our model. We therefore state the following.

\textbf{Proposition 2.} $g^*$ is a sequential-connection line network with even or odd active nodes if and only if it is a strict Nash network with the minimum number of links.

\textit{Proof.} Let $g^*$ be a sequential-connection line network, with an activation graph with even nodes and even $n$. For each $i$, the payoff is given by (4). Suppose that it is not a Nash network. That is, for at least one agent $i$, there exists a deviation $g'\in G$ such that
\[
\Pi_i(g')\neq
\begin{cases}
\displaystyle \sum_{h\in N} I_h-2, & \text{if } i \text{ is active},\\[6pt]
\displaystyle \sum_{h\in N} I_h-1, & \text{if } i \text{ is the last agent},\\[6pt]
\displaystyle \sum_{h\in N} I_h, & \text{if } i \text{ is passive}.
\end{cases}
\]

But the only way to achieve this is by reducing the number of links in the paths that carry information from all the other agents to $i$. If $i$ is passive, there is no strategy that yields a better payoff. If $i$ is active, she may reduce the links by one or by two. Since $g^*$ is a nonsequential-connection line network, there is only one agent $j$ such that
\[
g_{i,j}^*=1.
\]
There are three possible deviations for $g_i$.

First, suppose that $i$ decides to cut her link with agent $j+1$ but maintains her connection with $j-1$. With this, she saves one connection but loses her connection with agents $j+2,\ldots,n$. This implies that
\[
\sum I_i-1
<
\sum I_i.
\]
This is a contradiction.

Second, suppose that $i$ decides to cut her link with agent $j+1$ and, in turn, establish a link with agent $j+2$, maintaining that link with $j-1$. Then $i$ ceases to access the information supplied by agent $j+1$. But, by definition,
\[
v_{j+1}>c,
\]
so she would obtain a higher payoff by remaining connected. This is a contradiction.

Finally, consider the case in which $i$ chooses to disconnect and reduce her connection costs to zero. But in that case she accesses only her own information and ceases to access the information of the entire network. This is a contradiction. \hfill $\square$

If $g^*$ is a Nash equilibrium with the minimum number of links, then it constitutes a single component that includes all agents in $N$; otherwise, the information of agents who are not accessed would be lost for at least some other agent, while the reduction in link costs would not be sufficient to compensate for that loss. Recall that each $I_i$ is greater than the cost of one link. As shown previously, the minimum number of links that allows all agents to be connected is $n$.

\subsection*{III.4 Social Welfare}

According to Lemma 1 and Proposition 2, a stable result in the strategic interaction of agents is the sequential linear network with even or odd active nodes. It can be affirmed that it is stable because there are no incentives, once the structure of the circular network emerges, to cut links or form new ones, since the new configuration might not grant the same payoffs to the agents. The argument raises the question of the optimality of the result. That is, is there another configuration that can ensure better payoffs for the agents? Before answering this question negatively, we introduce two different notions of optimality that may be valuable to consider.

One represents the notion of social welfare ensured by the network. Formally, let
\[
W:G\to \mathbb{Z}
\]
be defined as
\[
W(g)=\sum_{i=1}^{n}\Pi_i(g)
\]
for
\[
g\in G.
\]
A network is said to be efficient if
\[
W(g)\geq W(g')
\]
for every
\[
g'\in G.
\]

On the other hand, we have the notion of Pareto optimality. A network $g$ is said to be Pareto optimal if there is no other network $g'$ such that, for every
\[
i\in N,
\]
\[
\Pi_i(g')\geq \Pi_i(g),
\]
and for at least one $i$,
\[
\Pi_i(g')>\Pi_i(g).
\]

We then reach the following result.

\textbf{Proposition 3.} A sequential line network with an activation graph of even or odd nodes that is strict Nash is both efficient and Pareto optimal.

\textit{Proof.} Recall that a strict Nash network $g^*$ sustains the maximum payoff for each agent:
\[
\Pi_i(g^*)=\sum_{h\in N} I_h(g)-2
\]
if $i$ is active,
\[
\Pi_i(g^*)=\sum_{h\in N} I_h(g)-1
\]
if $i$ is the last connected agent, and
\[
\Pi_i(g^*)=\sum_{h\in N} I_h(g)
\]
if $i$ is passive. Then,
\[
\Pi_i(g^*)\geq \Pi_i(g)
\]
for each
\[
i\in N
\]
and every
\[
g\in G.
\]
Thus, $g^*$ is optimal in the Pareto sense. By the same reasoning,
\[
W(g^*)=\sum_{i\in N}\Pi_i(g^*)
\geq
\sum_{i\in N}\Pi_i(g)=W(g)
\]
for every
\[
g\in G.
\]
That is, $g^*$ is efficient. \hfill $\square$

Table 5 below calculates the net social benefit of different network configurations under bidirectional information. All sequential line networks share the maximum benefit together with the centrally supported star network. However, as already mentioned, that configuration lacks a symmetry criterion that the linear configurations do possess, and therefore we believe it will be the configuration adopted by the agents.

\begin{longtable}{p{0.35\textwidth} p{0.55\textwidth}}
\caption{Payoff calculation by topology} \\
\toprule
\textbf{Topology} & \textbf{Net social benefit} \\
\midrule
\endfirsthead

\toprule
\textbf{Topology} & \textbf{Net social benefit} \\
\midrule
\endhead

1) Centrally supported star network &
Let
\[
S=\sum_{i\in N}I_i.
\]
Then
\[
\big(S-(n-1)\big)+(n-1)S
=
n(S-1)+1.
\]
\\[1.5em]

2) Sequential line networks & \\[0.5em]

2.a) Line network with ordered activation nodes &
\[
\sum_{i=1}^{n}\left(S-(n-i)\right)
=
\frac{1}{2}n(2S+1)-\frac{1}{2}n^2.
\]
\\[1.5em]

2.b) Line network with odd activation nodes, $n$ even &
\[
(S-1)+\left(\frac{n}{2}-1\right)(S-2)+\frac{n}{2}S
=
n(S-1)+1.
\]
\\[1.5em]

2.b) Line network with odd activation nodes, $n$ odd &
\[
(S-1)+\left(\frac{n-1}{2}\right)(S-2)+\frac{n+1}{2}S
=
n(S-1)+1.
\]
\\[1.5em]

3.b) Line network with even activation nodes, $n$ even &
\[
\left(\frac{n}{2}-1\right)(S-2)+\frac{n}{2}S+(S-1)
=
n(S-1)+1.
\]
\\[1.5em]

3.b) Line network with even activation nodes, $n$ odd &
\[
\left(\frac{n-1}{2}\right)(S-2)+\left(\frac{n+1}{2}\right)S
=
n(S-1)+1.
\]
\\

\bottomrule
\end{longtable}

\section*{Conclusion}

The restrictions imposed on benefits by the accumulative cost function greatly restrict the feasibility of finding a broad network as a strict Nash network. The optimal network cannot expand too much without incurring growing costs that affect agents' benefits. Under bidirectional information flow, some Nash networks are demarcated, with the least asymmetric one being the one that agents will end up adopting. This configuration is the sequential linear network with even or odd activation nodes, where intermediate agents make connections with their two immediate neighbors, cutting off the accumulation of costs of the previously defined function. This result also maximizes social welfare.

As future extensions of the paper, one may mention the interesting proposal of Berninghaus et al. (2004), taken up by Goyal (2004), suggesting the qualification of links so as to value the creation of links more than maintaining a passive attitude of allowing oneself to be connected. This suggestion could provide a broader topology as a strict Nash equilibrium solution for the mono- and bidirectional models.

\section*{Bibliography}

Berninghaus, S. K., M. Ott, and B. Vogt. 2004. \textit{On Networks and ``Stars'': Recent Results in Network Experiments}. University of Karlsruhe, mimeo.

Bala, V., and S. Goyal. 2000. ``A Noncooperative Model of Network Formation.'' \textit{Econometrica} 68: 1181--1229.  \url{https://doi.org/10.1111/1468-0262.00155}

Dutta, B., A. van den Nouweland, and S. Tijs. 1998. ``Link Formation in Cooperative Situations.'' \textit{International Journal of Game Theory} 27: 245--255. \url{https://doi.org/10.1007/s001820050070}

Falk, A., and M. Kosfeld. 2003. ``It's All About Connections: Evidence on Network Formation.'' IEW Working Paper 21, Universität Zurich.

Goyal, S. 2004. \textit{Strong and Weak Links}. Queen Mary University, London, mimeo.

Johnson, C., Gilles, R.P. (2003). Spatial Social Networks. In: Dutta, B., Jackson, M.O. (eds) \textit{Networks and Groups. Studies in Economic Design}. Springer, Berlin, Heidelberg. \url{https://doi.org/10.1007/978-3-540-24790-6_4}

Jackson, M., and A. Wolinsky. 1996. ``A Strategic Model of Social and Economic Networks.'' \textit{Journal of Economic Theory} 71: 44--74. DOI: 10.1006/jeth.1996.0108

Larrosa, J. M. C., and F. Tohm\'e. 2003. ``Formaci\'on de redes dirigidas circulares con costos de enlace compartidos.'' \textit{Estudios Econ\'omicos} XX(41), January--December: 27--48. \url{https://doi.org/10.52292/j.estudecon.2003.392} 

Qin, C. Z. 1996. ``Endogenous Formation of Cooperative Structures.'' \textit{Journal of Economic Theory} 69: 218--226. DOI: 10.1006/jeth.1996.0047

Slikker, M., R. Gilles, H. Norde, and S. Tijs. 2002. \textit{Directed Networks, Allocation Properties and Hierarchy Formation}. \textit{Mathematical Social Sciences} 49(1): 55-80. DOI: 10.1016/j.mathsocsci.2004.04.005

Slikker, M., and A. van den Nouweland. 2001. ``A One-Stage Model of Link Formation and Payoff Division.'' \textit{Games and Economic Behavior} 34: 153--175. DOI: 10.1006/game.1999.0785

\end{document}